\documentclass[aps,twocolumn]{revtex4}
\usepackage{graphicx}
\usepackage{dcolumn}
\usepackage{bm}
\usepackage{amsmath}

 \newcommand{\beq}[1]{\begin{equation}\label{#1}}
 \newcommand{\eeq}{\end{equation}}
 \newcommand{\bea}[1]{\begin{eqnarray}\label{#1}}
 \newcommand{\eea}{\end{eqnarray}}
 \newcommand\figcaption{\def\@captype{figure}\caption}
 \newcommand\tabcaption{\def\@captype{table}\caption}

\begin{document}

 \title{Enhanced nonreciprocal transmission through a saturable cubic-quintic nonlinear dimer defect}
 \author{Muhammad Abdul Wasay$^{1}$}
\email{wasay31@gmail.com}
\author{M. L. Lyra$^{2}$}
\email{marcelo@fis.ufal.br}
\author{B. S. Ham$^{1}$}
\email{bham@gist.ac.kr}

\affiliation{$^1$Center for Photon Information Processing, School of Electrical Engineering and Computer Science, Gwangju Institute of Science and Technology, Gwangju 61005, South Korea\\
$^2$Instituto de F\'{\i}sica, Universidade Federal de Alagoas, 57072-970 Macei\'o-AL, Brazil}
 \begin{abstract}
The transmission properties through a saturable cubic-quintic nonlinear defect attached to lateral linear chains is investigated. Particular attention is directed to the possible non-reciprocal diode-like transmission when the parity-symmetry of the defect is broken. Distinct cases of parity breaking are considered including asymmetric linear and nonlinear responses. The spectrum of the transmission coefficient is analytically computed and the influence of the degree of saturation analyzed in detail. The transmission of Gaussian wave-packets is also numerically investigated. Our results unveil that spectral regions with high transmission and enhanced diode-like operation can be achieved.

  \end{abstract}
 %\pacs{04.20.Cv,~03.65.Ta}

 \maketitle
 \smallskip

\section{Introduction}

The investigation of controlled transport of energy/mass has remained a hot area of research since a long time and it has recently gained considerable revival due to its direct implications in manufacturing technological devices for controlled transport. The diode is one of the basic devices which can directionally control
transport allowing it to occur mainly in a preferential direction. Besides the most traditional diodes that operate by rectifying electric current, there have been several demonstrations for the capability of achieving such non-reciprocal transport of heat flow~\cite{chang,sun,wang}, acoustic~\cite{li,boechler,yuan} and electromagnetic~\cite{gallo,fan,roy,lira} waves.

In linear systems with time-reversal symmetry, no symmetry breaking of transmission can be generated according to the reciprocity theorem~\cite{rayleigh,figotin,khanikaev}. Time-reversal symmetry can be broken in magneto-optical devices by an applied magnetic field to generate nonreciprocal transport of optical waves~\cite{lira}. An analog mechanism has been developed to produce non-reciprocal isolators of acoustic waves~\cite{fleury}. On the other hand, nonlinear processes can lead to asymmetric wave propagation without the need of an external magnetic field.
Recent experimental advances have demonstrated that  nonlinear optical lattices with alternating gain and loss atomic configurations are ideal realizations of non-Hermitian parity-time symmetric systems that can be explored to better understand the peculiar physical properties of these non-Hermitian systems in atomic settings\cite{nonH1,nonH2}.

By exploring to own nonlinear properties of the underlying medium, the nonreciprocal propagation of elastic and optical waves have been demonstrated. In particular, a pronounced rectifying factor for the scattering of harmonic waves was reported in a multilayered
system represented by a set of discrete nonlinear Schr\"{o}dinger equations with a
cubic nonlinearity restricted to take place just in
two nonlinear asymmetric layers~\cite{lepri}. Recently, the effect of a higher order quintic nonlinearity on the non-reciprocal transport has been investigated, unveiling that the combined effect of cubic and quintic terms is non-additive~\cite{wasay1}. The nonlinear dynamics of soliton formation in laser-induced optical gratings has been experimentally probed under the competing action of cubic and quintic nonlinear contributions\cite{comp1,comp2}, evidencing mutual transformations among droplet-like fundamental, dipole, and azimuthally modulated vortex solitons.

However, when very intense waves propagate in matter, it recovers its linear behavior but with a distinct group velocity as compared to that of
a low-intensity wave~\cite{agrawal}. This phenomenon is known as the saturation of the nonlinearity and is associated with the emergence of several non-trivial optical phenomena~\cite{agrawal,gatz,lyra1,zhong,nithyanandan,silva,lyra2,hu,cao,guzman,shi,samuelsen}. Within the context of non-reciprocal transmission through non-linear asymmetric layers, it has been shown that bistability can be enhanced by saturation of a cubic nonlinearity, thus favoring the nonreciprocal diode-like transmission. Also, saturation was shown to have opposite impacts of the rectifying action over short and long wavelength harmonic signals~\cite{assuncao}. Further, non-local  nonlinear responses were shown to generally act by reducing the diode action~\cite{wasay2}.

Motivated by the non-additivity of cubic and quintic nonlinear effects on the non-reciprocal transmission of harmonic waves and by the beneficial trend produced by saturation of the nonlinear responses, we will address to the question of how these aspects can be put together to enhance the diode-like scattering of waves by a nonlinear asymmetric dimer. Within a tight-binding approach, we will analytically compute the spectrum of transmission through a saturable nonlinear asymmetric dimer coupled to linear side-chains. We will explicitly consider the saturation of both cubic and quintic nonlinear contributions and explore distinct scenarios for breaking the
parity symmetry. We also numerically investigate the transmission of Gaussian incoming waves. In particular, we discuss the possibility of achieving large rectification action with high transmission in appropriated regions of the model's parameters space.

The paper is organized as follows: In section II, we introduce the model and develop the main formalism used to obtain the spectral transmission properties. In section III, we explore the saturation effect on the multistability and transmission for various(distinct) symmetry breaking scenarios. Section IV devotes to the analysis of the saturation effect on the rectification factor. In section V, we provide some numerical results concerning the non-reciprocal transmission on incoming Gaussian wave-packets.  Finally, in section VI, we summarize and draw our main conclusions.

%%%%%%%%%%%%%%%%%%%%%%%%%%%%%%%%%%%%%%%%%%%%%%%
\section{Saturable cubic-quintic dNLS chain}

In the present work, we study the stationary scattering solutions of plane waves in a discrete linear chain with a pair of saturable nonlinear defects.
The defects will be described as a saturable cubic-quintic discrete nonlinear Schr\"{o}dinger dimer.
The saturable cubic-quintic discrete nonlinear Schr\"{o}dinger (sCQDNLS) equation with a saturated nonlinearity is given by

\bea{}
i\dot{\psi}_n=V_n\psi_n-(\psi_{n+1}+\psi_{n-1})+\frac{\gamma_n|\psi_n|^2}{1+\mu_3|\psi_n|^2}\psi_n+\nonumber
\\
\frac{\nu_n|\psi_n|^4}{1+\mu_5|\psi_n|^4}\psi_n ,
\label{9th}
\eea
where $\mu_3$ and $\mu_5$ are the control parameters for the saturation of both the on-site cubic and quintic nonlinear responses, respectively. In what follows we will assume
$\mu_3=\mu_5=\mu$. $V_n$ is the potential at site $n$. The off-diagonal coefficient is chosen to be unity without loss of generality. The parameters $\gamma_n$ and $\nu_n$ are the local cubic and quintic nonlinear responses of the system, respectively, that are the main nonlinear contributions at low wave amplitudes.
 The stationary solutions of Eq. \eqref{9th} are given by

\bea{}
\psi_n(t)=\phi_n e^{-i\omega t} .
\eea

Using these in Eq. \eqref{9th}, we arrive at a stationary cq-DNLS equation

\bea{}
\omega\phi_n=-\phi_{n+1}-\phi_{n-1}+V_n\phi_n+\frac{\gamma_n|\phi_n|^2}{1+\mu|\phi_n|^2}\phi_n+\nonumber
\\
\frac{\nu_n|\phi_n|^4}{1+\mu|\phi_n|^4}\phi_n ,
\label{stationary}
\eea
which can be re-arranged in a backward iterative scheme \cite{lepri,tsironis} as follows:

\bea{}
\phi_{n-1}\!=\!-\phi_{n+1}\!+\! \left(\!V_n-\omega\!+\!\frac{\gamma_n|\phi_n|^2}{1+\mu|\phi_n|^2}\!+\!\frac{\nu_n|\phi_n|^4}{1+\mu|\phi_n|^4}\!\right)\! \phi_n ,
\label{10th}
\eea
where $\omega$ is the spatial frequency and $\phi_n$ is the complex mode amplitude at site $n$.
Eq.\eqref{10th} is a map for the amplitudes at site $n-1$ of the time-independent discrete nonlinear Schrodinger equation. This map is helpful for obtaining the transmission formulae, since Eq.\eqref{10th} provides the amplitude at site $n-1$ given  the amplitudes at site $n$ and $n+1$. Note that we are interested in examining the scattering by a nonlinear dimer, i.e., two sites carrying the nonlinear effects at the center of an infinite one dimensional chain. This implies that the full cq-DNLS in Eq.\eqref{9th} applies to just these two sites. Therefore, the last two terms in Eq.\eqref{9th} will not contribute when an input/output signal is away from the dimer. In that case, the transmission will be governed by just the linear part of Eq.\eqref{9th}.  Hence the input/output signal will not feel any nonlinear response from the lattice in that region and can propagate freely.

We will focus on the scattering properties of plane wave solutions of the following form

\bea{}
\phi_n=
\Bigg\{
  \begin{array}{c}
  R_0 e^{ikn}+Re^{-ikn}             \qquad n\leq 1
  \\ \\
  Te^{ikn}                          ~~\qquad\qquad\qquad n\geq 2 \\
  \end{array}
\eea
where, $R_0$, $R$ and $T$ are the amplitudes of incoming, reflected and transmitted waves, respectively.
Note that the nonlinear region in the lattice corresponds to the sites  $n=1,2$. Thus the site-dependent coefficients carrying nonlinear effects $\nu_n$ and $\gamma_n$ are absent in the linear part of the lattice. Therefore, the model corresponds to a nonlinear dimer connected to otherwise linear side chains. Further, we will consider  $V_n=0$ on both side chains.

The purpose of this work is to evaluate the capability of the saturated cq-DNLS dimer to produce efficient asymmetric scattering and hence to operate as a wave diode. The desired effect (asymmetric transmission) arises when one breaks the translational symmetry of the lattice \cite{lepri,assuncao,wasay1,wasay2}. This symmetry breaking combined with the nonlinearity leads to a nonreciprocal transmission of the input signal. In a one-dimensional setting, this can be done in different ways which we discuss in details in the following sections.

Using the backward transfer map, the transmission coefficient for a right-propagating wave ($k>0$, i.e., the wave is incident from the left of the dimer) is given by
\bea{}
t(k,|T|^2)=\frac{|T|^2}{|R_0|^2}=\left|\frac{e^{-ik}-e^{ik}}{(\Omega_1-e^{ik})(\Omega_2-e^{ik})-1}\right|^2 ,
\label{tc0}
\eea
where,
\bea{}
\Omega_2=V_2-\omega+\frac{\gamma_2|T|^2}{1+\mu|T|^2}+\frac{\nu_2|T|^4}{1+\mu|T|^4},
\label{delta2}
\eea
and
\bea{}
\Omega_1=V_1-\omega+~~~~~~~~~~~~~~~~~~~~~~~~~~~~~~~~~\nonumber
\\
\frac{\gamma_1|T|^2|\Omega_2-e^{ik}|^2}{1+\mu|T|^2|\Omega_2-e^{ik}|^2}+\frac{\nu_1|T|^4|\Omega_2-e^{ik}|^4}{1+\mu|T|^4|\Omega_2-e^{ik}|^4} .
\label{delta1}
\eea

When a harmonic wave is incident from the right of the nonlinear dimer ($k<0$), i.e., a left-propagating wave, the transmission coefficients can be computed in the same way simply by exchanging the subscripts 1 and 2, which amounts to flipping the lattice \cite{lepri,assuncao,wasay1,wasay2}.

\section{Effects of saturation on Multistability and transmission}

We are interested in exploring a nonreciprocal transmission of input signals which will lead us to the desired diode-like action. In particular, we focus on the influence of the simultaneous saturation of both cubic and quintic nonlinear responses. We begin by showing the typical relation between incoming and transmitted intensities for varying strengths of the saturation parameter $\mu$.
\begin{figure}[h!]
  \centering
  \includegraphics[scale=0.32]{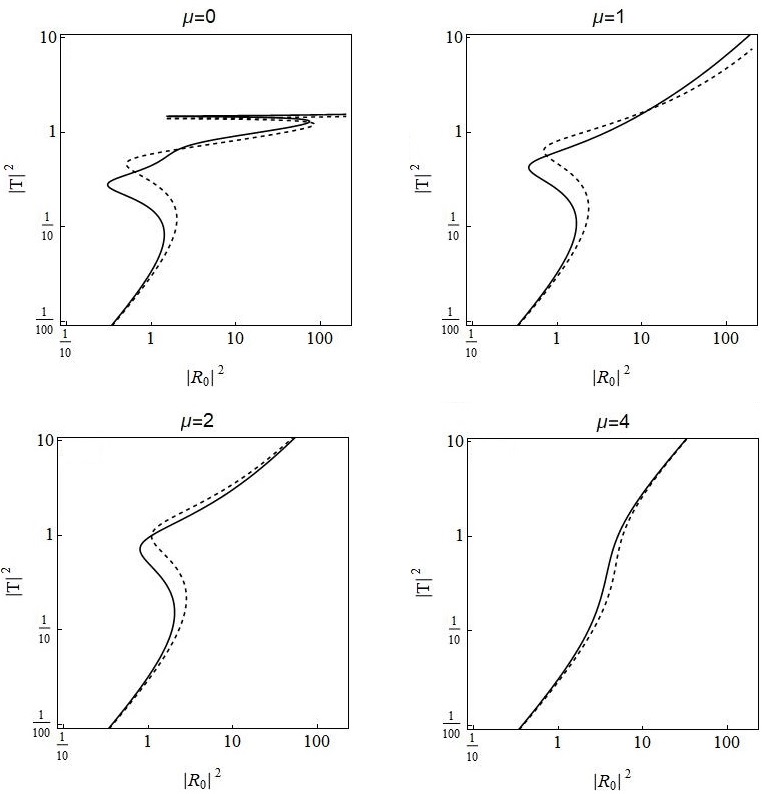}
  \caption{Relationship between incoming $|R_0|^2$ and transmitted $|T|^2$ wave intensities  for distinct values of saturation parameter $\mu$ starting from $\mu=0$ in the top left to $\mu=4$ in bottom right panel and with asymmetry between on-site potentials $\varepsilon_v=0.05$. Solid and dashed lines represent the right and left incidences. Here we considered an harmonic incoming mode with wave-number $k=0.1$. Model parameters are $V_0=-2.5$,$\gamma_1=\gamma_2=\gamma=1$, $\nu_1=\nu_2=\nu=0.5$,  $\varepsilon_v=0.05$}
  \label{v}
\end{figure}

%\begin{figure}[h]
%  \begin{minipage}[h]{0.477\linewidth}
%    \centering
%    \includegraphics[width=\linewidth]{v1.jpg}
%    %\caption{}
%  \end{minipage}
%  \hspace{0.2cm}
%  \begin{minipage}[h]{0.47\linewidth}
%    \centering
%    \includegraphics[width=\linewidth]{v2.jpg}
%    %\caption{ }
%  \end{minipage}
%  \begin{minipage}[h]{0.47\linewidth}
%    \centering
%    \includegraphics[width=\linewidth]{v3.jpg}
%    %\caption{ }
%  \end{minipage}
%  \hspace{0.2cm}
%  \begin{minipage}[h]{0.47\linewidth}
%    \centering
%    \includegraphics[width=\linewidth]{v4.jpg}
%    %\caption{ }
%  \end{minipage}
% \caption{Relationship between incoming $|R_0|^2$ and transmitted $|T|^2$ wave intensities  for distinct values of saturation parameter $\mu$ starting from $\mu=0$ in the top left to $\mu=4$ in bottom right panel and with asymmetry between on-site potentials $\varepsilon_v=0.05$. Solid and dashed lines represent the right and left incidences. Here we considered an harmonic incoming mode with wave-number $k=0.1$. Model parameters are $V_0=-2.5$,$\gamma_1=\gamma_2=\gamma=1$, $\nu_1=\nu_2=\nu=0.5$,  $\varepsilon_v=0.05$}
% \label{v}
%  \end{figure}

Figure \ref{v} is a plot for transmitted intensity $|T|^2$ as a function of incoming intensity $|R_0|^2$ for the nonlinear dimer with  $V_0=-2.5$ and with $\gamma=1$, $\nu=0.5$ representing the local (on-site) cubic and quintic nonlinear response of the lattice respectively, $\mu$ determines the extent to which we saturate these nonlinear responses.
Note that we have defined the asymmetry in on-site potentials $\varepsilon_v$ whose strength determines the extent to which the on-site potentials differ from each other  $V_1=V_0(1+\varepsilon_v)$ and $V_2=V_0(1-\varepsilon_v )$. In the case at hand we deal with a dimer and the difference between (two) on-site potentials is taken to be $\varepsilon_v=0.05$. Similarly, we will later utilize the asymmetry in other site dependent parameters with a relevant corresponding denotation. For no saturation ($\mu=0$, top left in Fig.\ref{v}), there are two bistability regions, i.e., regions where a single input intensity gets transmitted with different intensities representing distinct defect modes. The first window occurs at lower incident intensities and it is significantly broad compared to the second window occurring at higher incoming intensities (see \cite{wasay1} for details). When the saturation parameter $\mu$ increases, the multistability regions are suppressed, going from a single bistability window at intermediate values of $\mu$) to the usual single mode behavior at strongly saturable nonlinearities.

\begin{figure}[h!]
  \centering
  \includegraphics[scale=0.32]{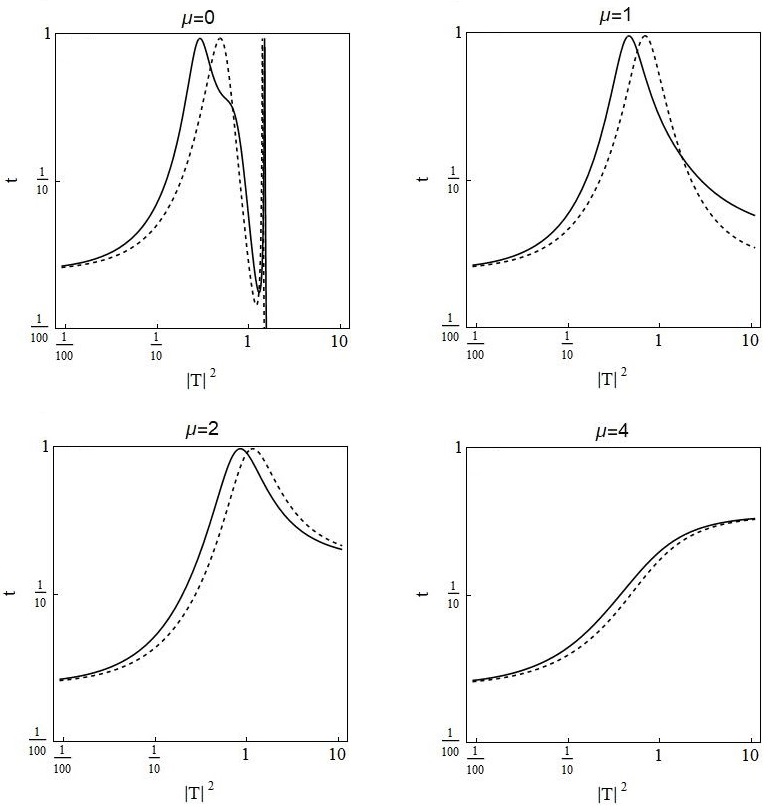}
  \caption{Transmission coefficient $t(k,|T|^2)$ as a function of transmitted wave intensity $|T|^2$ for distinct values of the saturation parameter $\mu$, and with $V_0=-2.5$, $\varepsilon_v=0.05$, $\gamma_{1,2}=1$ and $\nu_{1,2}=0.5.$. Solid curves correspond to the right incidence and dashed curves to the left incidence, with $k=0.1$.}
  \label{tvt}
\end{figure}

%\begin{figure}[h]
%  \begin{minipage}[h]{0.477\linewidth}
%    \centering
%    \includegraphics[width=\linewidth]{tvt1.jpg}
%    %\caption{}
%  \end{minipage}
%  \hspace{0.2cm}
%  \begin{minipage}[h]{0.47\linewidth}
%    \centering
%    \includegraphics[width=\linewidth]{tvt2.jpg}
%    %\caption{ }
%  \end{minipage}
%  \begin{minipage}[h]{0.47\linewidth}
%    \centering
%    \includegraphics[width=\linewidth]{tvt3.jpg}
%    %\caption{ }
%  \end{minipage}
%  \hspace{0.2cm}
%  \begin{minipage}[h]{0.47\linewidth}
%    \centering
%    \includegraphics[width=\linewidth]{tvt4.jpg}
%    %\caption{ }
%  \end{minipage}
% \caption{Transmission coefficient $t(k,|T|^2)$ as a function of transmitted wave intensity $|T|^2$ for distinct values of the saturation parameter $\mu$, and with $V_0=-2.5$, $\varepsilon_v=0.05$, $\gamma_{1,2}=1$ and $\nu_{1,2}=0.5.$. Solid curves correspond to the right incidence and dashed curves to the left incidence, with $k=0.1$. }
% \label{tvt}
%  \end{figure}

Figure \ref{tvt} shows the transmission coefficient $t(k,|T|^2)$ as a function of transmitted intensity $|T|^2$ for a sCQDNLS dimer with the same parameter values as in Fig.\ref{v}. The shift between the left and right-incident transmission peaks due to the asymmetry in on-site potentials $\varepsilon_v$ is apparent. It is also important to mention that the transmission coefficient is comparatively larger in the ranges of intensity which lead to a significantly asymmetric transmission. The asymmetry between the left and right incidence transmission is suppressed when the nonlinear dimer is strongly saturable.

  \begin{figure}[h]
  \centering
  \includegraphics[scale=0.35]{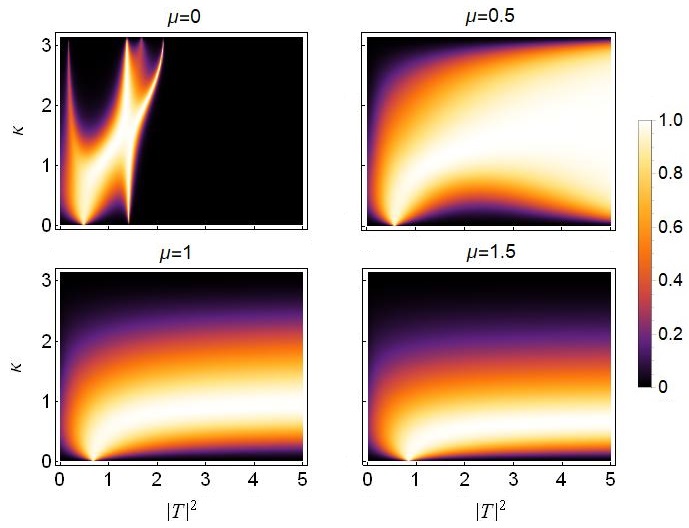}
  \caption{Density plot of transmission coefficient as a function of $k$ and $|T|^2$: From $\mu=0$ in top left to $\mu=1.5$ in bottom right and with $\varepsilon_v=0.05$. Other parameters are the same as those used in the previous figures.}
  \label{dpd}
\end{figure}

Fig.\ref{dpd} is the transmission coefficient density plot for the sCQDNLS dimer with asymmetric on-site potentials and with the parameter values for the on-site nonlinear responses being the same as in the previous figures. The transmission increases with increasing saturation from $\mu=0$ to $\mu=0.5$ (see the plots in the top panel in Fig.\ref{dpd}), then the transmission gradually decreases when we further increase the saturation from $\mu=0.5$ to $\mu=1.5$ (see the plots in the bottom panel in Fig.\ref{dpd}).

\begin{figure}[h]
  \centering
  \includegraphics[scale=0.37]{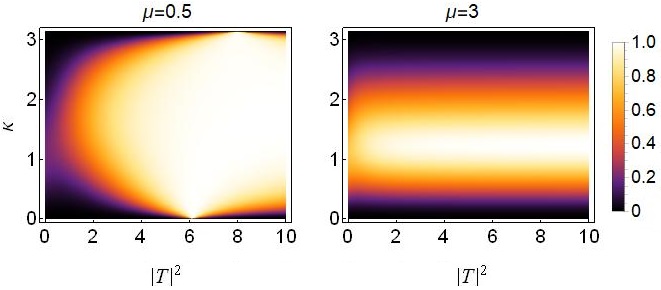}
  \caption{Density plot of transmission coefficient as a function of $k$ and $|T|^2$ for varying saturation. Left Plot: $\mu=0.5$, Right Plot: $\mu=3$. Site dependent parameter values are $V_{1,2}=V_0(1\pm\varepsilon_v)$ with $V_0=-2.5$, $\varepsilon_v=0.8$; $\gamma_{1,2}=\gamma_0(1\pm\varepsilon_c)$ with $\gamma_0=1$, $\varepsilon_c=0.5$; $\nu_{1,2}=\nu_0(1\pm\varepsilon_q)$ with $\nu_0=0.5$, $\varepsilon_q=1.2$. }
  \label{dpd1}
\end{figure}

When the lattice's symmetry is broken threefold with all three site dependent parameters chosen to be different simultaneously at each dimer site (different on-site potential, cubic and quintic response), it is possible to maintain higher transmission at increased levels of asymmetry and saturation. Figure \ref{dpd1} is produced to depict this case with the parameter values chosen as described in the caption. The three-fold symmetry breaking parameters $\varepsilon_v$, $\varepsilon_c$ and $\varepsilon_q$ are non-zero simultaneously, where $\varepsilon_v$ describes the asymmetry in on-site potentials, $\varepsilon_c$ and $\varepsilon_q$ describe the asymmetry in cubic and quintic response, respectively. In the regime of high saturation and 3-fold asymmetry, high transmission occurs at intermediate wave-numbers $k\approx0.8\sim1.8$. Note however that in this regime, there is no transmission for very small wave-numbers $k\approx0\sim0.2$ or very large wave-numbers $k\approx2.7\sim\pi$. It is important to stress that higher transmission at higher asymmetry is a welcome feature with regard to a diode-like transmission. In previous works,  higher asymmetry renders a better diode effect but usually with an overall lower transmission\cite{lepri,wasay1,wasay2}. The present result shows that the simultaneous asymmetry on the potential and cubic and quintic nonlinear contributions can overcome the problem of lower transmission at better diode-like modes.

\section{Distinct asymmetry scenarios}

As mentioned before, there are three distinct ways to break the translational symmetry of the underlying one-dimensional sCQDNLS lattice system in order to support the nonreciprocal transmission. It is of interest to study possible implications of saturating the nonlinear responses of the lattice on transmission of input signals under various distinct possibilities to induce asymmetry. The sCQDNLS dimer model has two local nonlinear responses. In this section, we will report on the effects of saturation on the nonreciprocal transmission under three different possible parity breaking schemes.  The three different cases correspond to the three distinct choices of on-site parameters asymmetries. For instance, one could choose either of the following
 \\(i):  ~different on-site potentials for the two dimer sites.
 \\ (ii): different on-site cubic nonlinear response.
 \\ (iii):\! different on-site quintic nonlinear response.

Additionally, one can also have all three factors acting simultaneously, which would amount to a higher degree of broken symmetry, i.e., a threefold-symmetry breaking or threefold-asymmetry case, which we will also discuss in detail in the upcoming sub-sections.

 \subsection{Different on-site potentials}

\begin{figure}[h]
  \centering
  \includegraphics[scale=0.37]{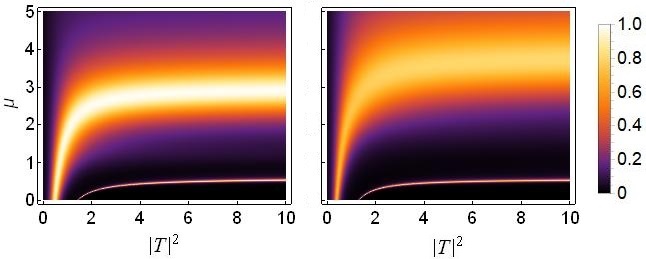}
  \caption{Density plot of transmission coefficient as function of $\mu$ and $|T|^2$ for $k=0.1$: Left: $\varepsilon_v=0.05$ Right: $\varepsilon_v=0.2$. Both plots are produced for $\varepsilon_c=\varepsilon_q=0$, $\gamma_0=1$, $\nu_0=0.5$ and $V_0=-2.5$.}
  \label{tcu1}
\end{figure}

 This is the first of three possibilities to break the translational symmetry of the one dimensional chain under consideration. Figure \ref{tcu1} presents the transmission coefficient as a function of saturation $\mu$ and the transmitted intensity $|T|^2$.

\begin{figure}[h]
  \centering
  \includegraphics[scale=0.4]{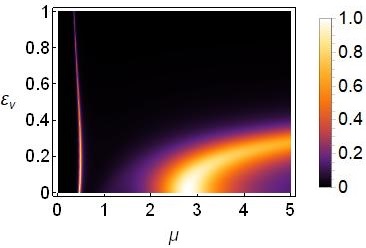}
  \caption{Density plot of transmission coefficient as function of $\mu$ on horizontal axis versus $\varepsilon_v$ on vertical axis. The plot is produced for $k=0.1$, $|T|^2=5$, $\varepsilon_q=\varepsilon_c=0$, $\gamma_0=1$, $\nu_0=0.5$ and $V_0=-2.5$.}
  \label{tcuc1}
\end{figure}

As shown in Fig.\ref{tcu1}, bulk of the transmission occurs at saturation $\mu=2\sim3$. For smaller saturation, $\mu<0.5$, the transmission is very small. To reinforce this conclusion, we plot the transmission coefficient as function of saturation parameter $\mu$ and asymmetry in on-site potentials $\varepsilon_v$ in Fig.\ref{tcuc1} which confirms that the transmission decreases at higher asymmetry, i.e., $\varepsilon_v>0.2$ and is entirely suppressed at $\varepsilon_v\geq0.4$.

\subsection{Different on-site cubic nonlinearities}

\begin{figure}[h]
  \centering
  \includegraphics[scale=0.37]{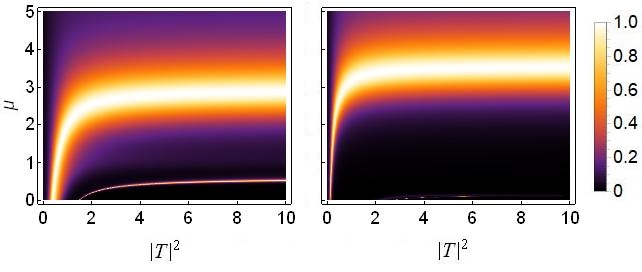}
  \caption{Density plot of transmission coefficient as function of $\mu$ and $|T|^2$ for $k=0.1$. The cubic response $\gamma_{1,2}=\gamma_0(1\pm\varepsilon_c)$ with $\gamma_0=1$: Left: $\varepsilon_c=0.05$ Right: $\varepsilon_c=1.5$. Both plots are produced for $\varepsilon_v=\varepsilon_q=0$, $\nu_0=0.5$ and $V_0=-2.5$.}
  \label{tcu2}
\end{figure}

As discussed above, it is also possible to break the lattice's symmetry by choosing different cubic nonlinearities for the two sites under consideration i.e., $\gamma_1\neq\gamma_2$, the resulting effects on transmission are depicted in Fig.\ref{tcu2}.

Comparing the plots in Fig.\ref{tcu1} and Fig.\ref{tcu2}, it is evident that the first (left) plot from both figures are quite similar. Hence, the transmission remains almost the same in both cases for small asymmetry. However, the plot on the right is very different in the two figures. It shows that the trend of diminishing transmission with increasing asymmetry in on-site potentials ($\varepsilon_v\neq0$) does not hold in the case when asymmetry is between the cubic nonlinear responses ($\varepsilon_c\neq0$). It is clear from Fig.\ref{tcu2} (right) that even at a sufficiently larger asymmetry ($\varepsilon_c$), the system still allows significant transmission roughly for the saturation values of $\mu=2.5\sim3.8$.

\begin{figure}[h]
  \centering
  \includegraphics[scale=0.4]{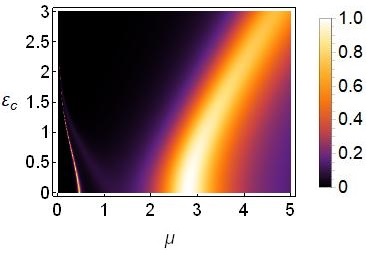}
  \caption{Density plot of transmission coefficient as function of $\mu$ on horizontal axis versus $\varepsilon_c$ on vertical axis for $k=0.1$, $|T|^2=5$, $\varepsilon_v=\varepsilon_q=0$, $\gamma_0=1$, $\nu_0=0.5$ and $V_0=-2.5$. }
  \label{tcuc2}
\end{figure}

The corresponding plot of transmission coefficient as function of $\mu$ and $\varepsilon_c$ is plotted in Fig.\ref{tcuc2}, which confirms our analysis that a significant transmission indeed survives for large enough $\varepsilon_c$ (Notice the scale on vertical axes in Fig.\ref{tcuc1} and Fig.\ref{tcuc2}).

\subsection{Different on-site quintic nonlinearities}

\begin{figure}[h]
  \centering
  \includegraphics[scale=0.37]{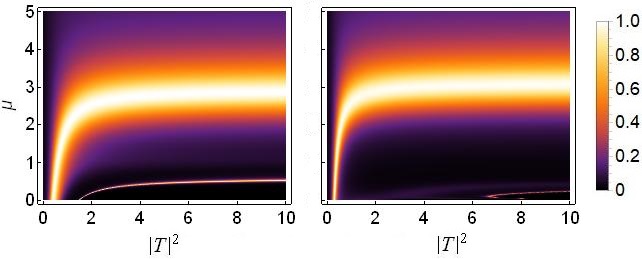}
  \caption{Density plot of transmission coefficient as function of $\mu$ and $|T|^2$ for $k=0.1$: The quintic response $\nu_{1,2}=\nu_0(1\pm\varepsilon_q)$ with $\nu_0=0.5$. Left: $\varepsilon_q=0.05$ Right: $\varepsilon_q=1.5$. Both plots are produced for $\varepsilon_v=\varepsilon_c=0$, $\gamma_0=1$ and $V_0=-2.5$.}
  \label{tcu3}
\end{figure}

\begin{figure}[h]
  \centering
  \includegraphics[scale=0.4]{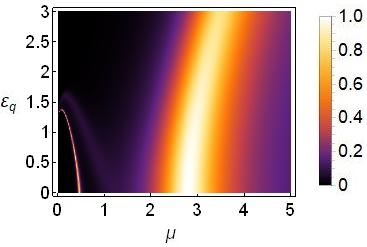}
  \caption{Density plot of transmission coefficient as function of $\mu$ on horizontal axis versus $\varepsilon_q$ on vertical axis for $k=0.1$, $|T|^2=5$, $\varepsilon_v=\varepsilon_c=0$, $\gamma_0=1$, $\nu_0=0.5$ and $V_0=-2.5$.}
  \label{tcuc3}
\end{figure}

Now we consider the effect of saturation on transmission when the lattice symmetry is broken by different on-site quintic nonlinearity i.e., $\nu_1\neq\nu_2$. Figure \ref{tcu3} shows the transmission in the $\mu$ versus $|T|^2$ plane for varying asymmetry $\varepsilon_q$ in the quintic response. The corresponding plot for $\mu$ versus $\varepsilon_q$ is given in Fig.\ref{tcuc3}

Figure \ref{tcuc3} confirms that when the lattice symmetry is broken by choosing different quintic responses ($\varepsilon_q\neq0$), transmission of the input signal remains pronounced for even higher asymmetry compared with the cases discussed before.

\subsection{Threefold symmetry breaking}

From Fig.\ref{tcuc1} we learn that for those parameter regimes, bulk of the transmission occurs for asymmetry $\varepsilon_v\leq 0.2$, and for saturation $\mu\approx2.5\sim3.5$. The cases of symmetry breaking by means of nonlinearity Fig.\ref{tcuc2} and Fig.\ref{tcuc3} suggest that it is possible to achieve higher transmission regimes for increased asymmetry levels which could be interesting to explore in the context of a diode-like transmission because, as discussed earlier in this paper and other studies \cite{lepri,assuncao,wasay1,wasay2}, higher asymmetry renders a better diode-like action but with reduced overall transmission.

Below we show the density plot of transmission coefficient as a function of $|T|^2$ and $\varepsilon_v$ (left) and a corresponding plot of transmission coefficient as a function of $\varepsilon_v$ and $\mu$ (right), for the case of a threefold broken symmetry of the one-dimensional sCQDNLS lattice.

\begin{figure}[h]
  \centering
  \includegraphics[scale=0.37]{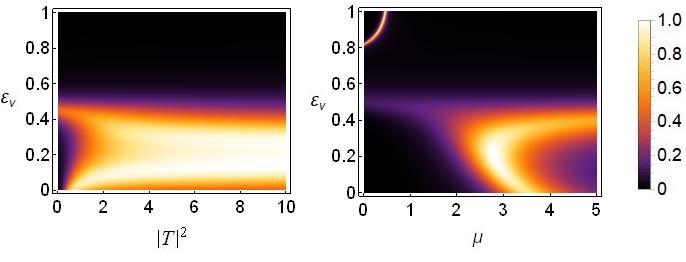}
  \caption{Density plot of transmission. Left: Transmission coefficient as function of $|T|^2$ and $\varepsilon_v$ for saturation $\mu=3$. Right: Transmission coefficient as a function of $\mu$ and $\varepsilon_v$ produced for $|T|^2=5$. Both plots are for the threefold broken symmetry case with $\varepsilon_c=\varepsilon_q=1$ and $k=0.1$ and all other parameters as before.}
  \label{tcuc4}
\end{figure}

It is clear from Fig.\ref{tcuc4} that the threefold broken symmetry can help lift the transmission regimes up towards higher asymmetry. However, bulk of the transmission still occurs for saturation values around $\mu=2.5\sim3.5$.

In summary, we can conclude from the results presented in Fig.\ref{tcuc1}, Fig.\ref{tcuc2}, Fig.\ref{tcuc3} and Fig.\ref{tcuc4} that the transmission is more susceptible to asymmetry when the broken translational symmetry (or asymmetry leading to nonreciprocal transmission) corresponds to different on-site potentials (i.e., $\varepsilon_v$ in our notation), as compared to either of the other two types of asymmetries (i.e., $\varepsilon_c$ or $\varepsilon_q$ in our notation).

\section{Effects of saturation on rectification}

To extend our analysis on the diode-like transmission through the nonlinear dimer, we will present some results for the so-called rectifying factor in this section, which will allow us to identify regions with maximal asymmetry in the transmission. The rectifying factor is defined as \cite{lepri}
\bea{}
\mathcal{R}=\frac{t_R(k,|T|^2)-t_L(k,|T|^2)}{t_R(k,|T|^2)+t_L(k,|T|^2)} ,
\label{11th}
\eea
where $t_L$ and $t_R$ are the transmission coefficients of the waves coming from the left and right of the nonlinear dimer, respectively. The rectifying factor $\mathcal{R}$ has a maximum value of $\pm1$, which corresponds to perfect diode-like action.
In what follows we will present the rectifying action for the cases when (i) asymmetry in the lattice is due to different on-site potentials, i.e., $V_1\neq V_2$ (ii) asymmetry due to different on-site cubic nonlinear response, i.e., $\gamma_1\neq\gamma_2$ , (iii) asymmetry due to different on-site quintic nonlinear response, i.e., $\nu_1\neq\nu_2$. We will highlight the effect of saturation on rectification for these three cases and also for the threefold-asymmetry case.

\subsection{Asymmetry due to different on-site potentials}

\begin{figure}[h]
  \centering
  \includegraphics[scale=0.36]{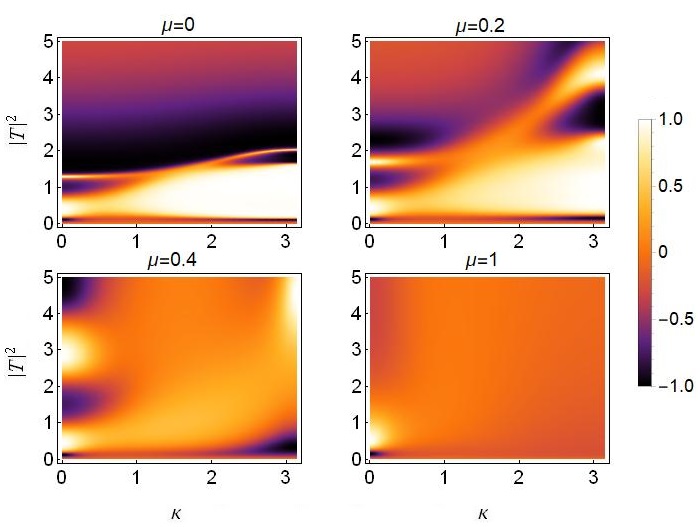}
  \caption{Density plot of rectifying factor for increasing saturation from $\mu=0$ in top left to $\mu=1$ in bottom right panel. Plot is produced for different on-site potentials $V_{1,2}=V_0(1\pm\varepsilon_v)$ with asymmetry $\varepsilon_v=0.2$ and $V_0=-2.5$ and with $\varepsilon_c=\varepsilon_q=0$, $\gamma_0=1$, $\nu_0=0.5$.}
  \label{rfd}
\end{figure}

Rectifying action with different on-site potentials under the effect of varying saturation is plotted in Fig.\ref{rfd}. Note that  the asymmetry between on-site potentials remains fixed at $\varepsilon_v=0.2$, cubic response at $\gamma_1=\gamma_2=1$ and quintic response at $\nu_1=\nu_2=0.5$. The transmission becomes more symmetric as the saturation coefficient increases, specially for $\mu>0.4$.

For the cases of $\mu\leq0.4$, a diode-like action is apparent, with transmission of both right-propagating (brighter) and left-propagating (darker) branches. This action seems to be more pronounced for input waves with larger wave-numbers for saturation up to $\mu=0.2$. For saturation $\mu>0.2$ it mostly favors the asymmetric transmission of  input waves with small wave-numbers, with the exception of a couple of faint branches exhibiting some diode-like action at large wave-numbers $k\sim\pi$ at $\mu=0.4$. Let us denote this kind of action as ``\textit{positive diode-action}", for reasons which will become clear in the upcoming sections.

%\begin{figure}[h]
%  \centering
%  % Requires \usepackage{graphicx}
%  \includegraphics[scale=0.47]{rfd1.jpg}
%  \caption{Rectifying factor for increasing saturation from $\mu=0$ in top left to $\mu=1$ in bottom right. For asymmetric on-site potentials with asymmetry $\varepsilon_v=0.8$}
%  \label{rfd1}
%\end{figure}

%Comparing the two plots in Fig.\ref{rfd} and Fig.\ref{rfd1}, it is evident that the diode-action seems to be better with increased $\varepsilon_v$, but only for smaller saturation i.e., upto $\mu\sim0.2$. As the saturation increases, the nonlinear effects arising from the two on-site nonlinear sources $\gamma$ and $\nu$, become less effective, as is well-known that higher saturation renders the model more and more linear [References].
Furthermore, it is also apparent that the diode-action is more pronounced for larger wave-numbers, particularly for smaller saturation values as evidenced in the top panel of Fig.\ref{rfd}. Note, however, that even though a diode-like effect increases with increasing asymmetry, the overall transmission reduces significantly, as noted before in \cite{lepri,wasay1,wasay2}.

\subsection{Asymmetry due to different on-site cubic nonlinear responses}

\begin{figure}[h!]
  \centering
  \includegraphics[scale=0.36]{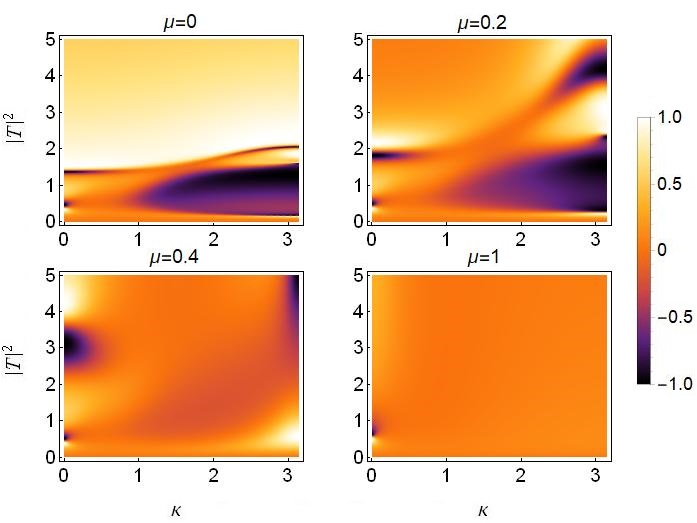}
  \caption{Density plot of rectifying factor for increasing saturation from $\mu=0$ in top left to $\mu=1$ in bottom right. For asymmetric on-site cubic nonlinear response $\gamma_{1,2}=\gamma_0(1\pm\varepsilon_c)$ with $\gamma_0=1$ and $\varepsilon_c=0.2$. Plots are produced for $\varepsilon_v=\varepsilon_q=0$, $\nu_0=0.5$, $V_0=-2.5$.}
  \label{rfd(gamma)}
\end{figure}

Rectifying action with different on-site cubic nonlinearity and under the effect of varying saturation is plotted in Fig.\ref{rfd(gamma)}. The nonlinearity is taken to be different on the two dimer sites to induce the required asymmetry in the system. The asymmetry in cubic response is fixed at $\varepsilon_c=0.2$. Other site-dependent parameters as $\nu_1=0.5=\nu_2$ and $V_1=-2.5=V_2$. The wave fills the nonlinear responses as $\gamma_{1,2}=\gamma(1\pm\varepsilon_c)$ at the first and second dimer site respectively. Comparing results from Fig.\ref{rfd} and Fig.\ref{rfd(gamma)}, it becomes clear that the diode-action is reversed for this case.

\subsection{Asymmetry due to different on-site quintic nonlinear responses}

\begin{figure}[h!]
  \centering
  \includegraphics[scale=0.36]{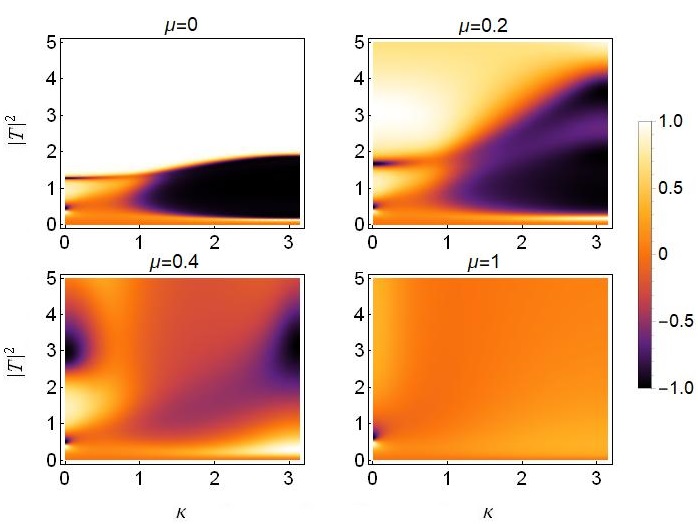}
  \caption{Density plot of rectifying factor for distinct increasing saturation from $\mu=0$ in top left to $\mu=1$ in bottom right. For asymmetric on-site quintic nonlinear response $\nu_{1,2}=\nu_0(1\pm\varepsilon_q)$ with $\nu_0=0.5$ and $\varepsilon_q=0.8$. Plots are produced for $\varepsilon_v=\varepsilon_c=0$, $\gamma_0=1$, $V_0=-2.5$.}
  \label{rfd(nu)}
\end{figure}

It is important to mention that for the case of asymmetric on-site quintic nonlinearity with $\nu=0.5$, $\gamma=1$, $\varepsilon_q=0.2$, $\nu_{1,2}=\nu_0(1\pm\varepsilon_q)$, we get the same rectifying action as that in the case of asymmetric cubic nonlinearity shown in Fig.\ref{rfd(gamma)}. Therefore, we can conclude that, for a specific value of asymmetry between cubic response and quintic response, the rectifying action is similar. So it is possible to achieve a similar diode-like action for either of the cases discussed above with same site-dependent parameter strengths. This trend can carry on for various strengths of saturation up to $\mu\sim1$. Therefore, due to the above reasoning and the fact that it is possible to maintain pronounced transmission at higher asymmetry, we present the rectifying plots for this case with a higher value of asymmetry between on-site quintic nonlinearity $\varepsilon_q=0.8$, as shown in Fig.\ref{rfd(nu)}.

Note that Fig.\ref{rfd(nu)} is produced for a higher asymmetry value i.e., $\varepsilon_q=0.8$ and the corresponding diode-like action is visible, which remains pronounced for values of saturation up to $\mu\sim0.4$. In the limit of higher saturating, the pattern seems to align with the asymmetric cubic case discussed in Fig.\ref{rfd(gamma)}.

It is also important to note that the diode-like action is reversed in comparison to the case of asymmetric on-site potentials, (see Fig.\ref{rfd}). The left-propagating (dark region) waves with large wave-numbers are transmitted while right-propagating (clear region) waves with smaller wave-numbers get through. This pattern carries on for all saturation strengths. Let's denote this kind of action as \textit{negative diode-action}, which will be a handy denotation in the upcoming section.

\subsection{Threefold asymmetry}

The threefold asymmetry case provides us with a variety of control parameters to be able to manipulate for the desired kind of diode-action. The asymmetry parameters related to nonlinearity (cubic and quintic) i.e., $\varepsilon_q$, $\varepsilon_c$ compete with the asymmetry parameter $\varepsilon_v$ (asymmetry in on-site potential), for the kind of diode-like action, discussed in previous subsections. Both $\varepsilon_q$ and $\varepsilon_c$ favor a \textit{negative diode-action}, while strengthening $\varepsilon_v$ helps produce a  \textit{positive diode-action}. Moreover, we also report that the positive action is more susceptible to $\varepsilon_v$ as compared to $\varepsilon_q$ and $\varepsilon_c$, i.e., comparatively larger values of $\varepsilon_q$ and $\varepsilon_c$ are required to suppress the positive action produced by a smaller $\varepsilon_v$ value. This conclusion also confirms our results from section-IV.

\begin{figure}[h!]
  \centering
  \includegraphics[scale=0.35]{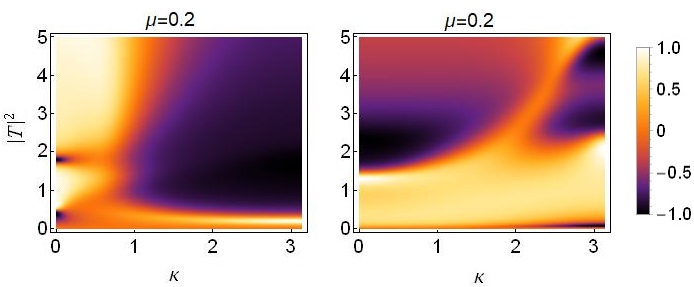}
  \caption{Density plot of rectifying factor for distinct (nonzero) values of three-fold asymmetries. Left: $\varepsilon_q=\varepsilon_c=0.7$ and $\varepsilon_v=0.2$. Right: $\varepsilon_q=\varepsilon_c=0.5$ and $\varepsilon_v=0.8$. All other parameters as before. }
  \label{rfd(3fold)}
\end{figure}

Plots in Fig.\ref{rfd(3fold)} are produced for a fixed value of saturation $\mu=0.2$. The branches exhibiting the \textit{positive} and \textit{negative} diode-action are apparent which clearly demonstrate the reverse action.

%\textcolor[rgb]{1.00,0.00,0.00}{\textit{Do you think it is also necessary to discuss the threefold case in detail and present the results for varying $\mu$? The results for the threefold case are almost exactly the same as asymmetric cubic and quintic cases discussed above. }}

\section{Propagation of a Gaussian wave-packet}

\begin{figure}[h!]
  \centering
  \includegraphics[scale=0.28]{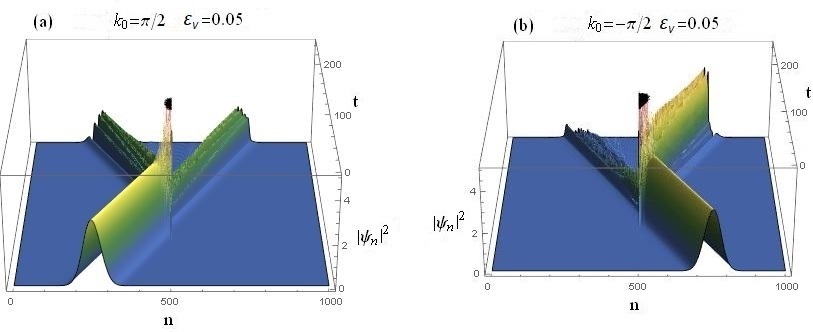}
  \caption{Time-dependent dynamics of a Gaussian wave-packet scattered by a sCQDNLS dimer placed at the center, with $V_0=-2.5$, $\gamma_0=1$, $\mu=0.1$, $\nu_0=0.5$, $\varepsilon_v=0.05$, $\varepsilon_q=\varepsilon_c=0$. (a): Right incidence. (b): Left incidence.}
  \label{GWP}
\end{figure}

It is instructive to consider the implications of the above considerations on the transmission of a Gaussian wave-packet. In this section, we consider the time-dependent dynamics of a Gaussian wave-packet propagating in the sCQDNLS lattice. A Gaussian wave-packet is taken as the initial condition \cite{lepri,wasay1}

\bea{}
\phi_n(0)= B~ \textmd{exp}\left[-\frac{\left(n-n_0\right)^2}{\sigma^2}+ik_0n\right] ,
\label{gwp}
\eea
where $B$ is the amplitude and $\sigma$ the width of the initial wave-packet chosen to be $\sigma=56$ in all the results reported in this section. With the sCQDNLS dimer situated at the center, the scattering of an initial input signal constituting of a Gaussian wave-packet  is given in Fig.\ref{GWP} for asymmetric on-site potential. The case for both left and right incidences are shown.

%\begin{figure}[h]
%  \begin{minipage}[h]{0.477\linewidth}
%    \centering
%    \includegraphics[width=\linewidth]{GWPR.jpg}
%    %\caption{}
%  \end{minipage}
%  \hspace{0.1cm}
%  \begin{minipage}[h]{0.477\linewidth}
%    \centering
%    \includegraphics[width=\linewidth]{GWPL.jpg}
%    %\caption{ }
%  \end{minipage}
% \caption{Time-dependent dynamics of a Gaussian wave-packet scattered by a sCQDNLS dimer placed at the center, with $V_0=-2.5$, $\gamma_0=1$, $\mu=0.1$, $\nu_0=0.5$, $\varepsilon_v=0.05$, $\varepsilon_q=\varepsilon_c=0$. (a): Right incidence. (b): Left incidence.}
% \label{GWP}
%  \end{figure}

  It is evident that, due to the broken parity symmetry in this system, right incidence gets a higher transmission (see Fig.\ref{GWP}(a)) as compared to the left incidence in Fig.\ref{GWP}(b). The corresponding transmission coefficients are $t_{k>0}=0.576647$ and $t_{k<0}=0.157794$. Note that the wave-packet scattering plot in Fig.\ref{GWP} is produced for asymmetric on-site potentials with the asymmetry $\varepsilon_v=0.05$, saturation $\mu=0.1$, cubic and quintic nonlinearities at $\gamma=1$ and $\nu=0.5$, respectively.
\begin{figure}[h!]
  \centering
  \includegraphics[scale=0.28]{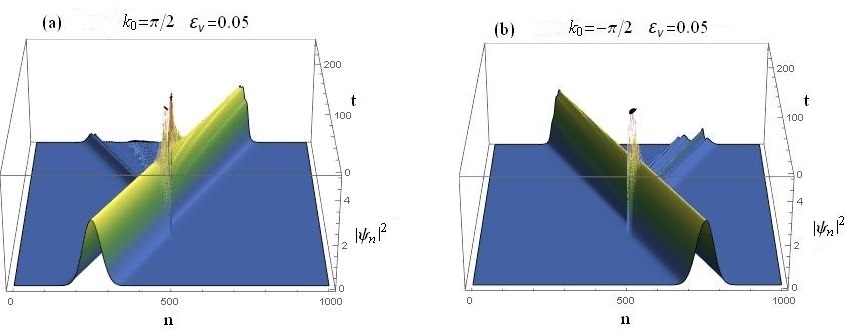}
  \caption{Time-dependent dynamics of a Gaussian wave-packet at a higher value of saturation $\mu=0.5$ and all other parameters as before. (a): Right incidence. (b): Left incidence.}
  \label{GWP1}
\end{figure}

%\begin{figure}[h]
%  \begin{minipage}[h]{0.477\linewidth}
%    \centering
%    \includegraphics[width=\linewidth]{GWPR1.jpg}
%    %\caption{}
%  \end{minipage}
%  \hspace{0.1cm}
%  \begin{minipage}[h]{0.477\linewidth}
%    \centering
%    \includegraphics[width=\linewidth]{GWPL1.jpg}
%    %\caption{ }
%  \end{minipage}
% \caption{Time-dependent dynamics of a Gaussian wave-packet at a higher value of saturation $\mu=0.5$ and all other parameters as before. (a): Right incidence. (b): Left incidence.}
% \label{GWP1}
%  \end{figure}

  One can conclude from Fig.\ref{GWP1} that the wave-packet seems to have a much improved transmission when we saturate both the nonlinearities to a higher level $\mu=0.5$, although with a degraded rectification action. The incident wave-packet also seems to have maintained its shape after transmission through the two nonlinear layers i.e., the dimer. The corresponding transmission coefficients are  $t_{k>0}=0.861496$ and $t_{k<0}=0.762887$.

  \subsection{Threefold symmetry breaking}

  Here, we present the results of the wave-packet dynamics when the underlying lattice system exhibits a threefold symmetry breaking. i.e., all three site-dependent parameters $V$,$\gamma$ and $\nu$ are different for the two dimer sites, and all other parameters are the same as in the previous cases of Fig.\ref{GWP} and Fig.\ref{GWP1}.

  \begin{figure}[h!]
  \centering
  \includegraphics[scale=0.28]{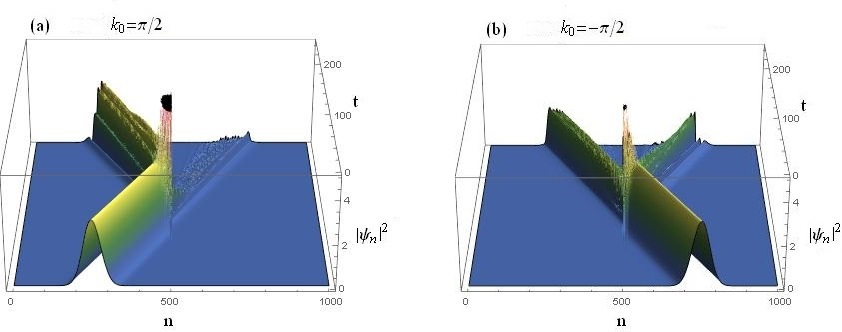}
  \caption{Time-dependent dynamics of a Gaussian wave-packet at $\mu=0.1$ for the case of threefold symmetry breaking with $\varepsilon_v=0.05$, $\varepsilon_c=0.2$, $\varepsilon_q=0.2$, and all other parameters as before. (a): Right incidence. (b): Left incidence.}
  \label{GWP3}
\end{figure}

  %\begin{figure}[h]
%  \begin{minipage}[h]{0.477\linewidth}
%    \centering
%    \includegraphics[width=\linewidth]{GWPR3.jpg}
%    %\caption{}
%  \end{minipage}
%  \hspace{0.1cm}
%  \begin{minipage}[h]{0.477\linewidth}
%    \centering
%    \includegraphics[width=\linewidth]{GWPL3.jpg}
%    %\caption{ }
%  \end{minipage}
% \caption{Time-dependent dynamics of a Gaussian wave-packet at $\mu=0.1$ for the case of threefold symmetry breaking with $\varepsilon_v=0.05$, $\varepsilon_c=0.2$, $\varepsilon_q=0.2$, and all other parameters as before. (a): Right incidence. (b): Left incidence.}
% \label{GWP3}
%  \end{figure}

  The wave-packet transmission coefficients for the case discussed in Fig.\ref{GWP3} are $t_{k>0}=0.133193$ and $t_{k<0}=0.669277$ which, together with Fig.\ref{GWP1}, suggests that the left propagating wave-packets have a significantly higher transmission rate as compared to the right propagating ones. Hence the trend is reversed under the threefold symmetry breaking which again confirms our analysis on the \textit{reverse} diode-action discussed in detail in the previous section. It occurs for the case with higher saturation with roughly the same difference between the left and right transmission coefficients as in case of Fig.\ref{GWP1}.

  \section{Summary and conclusion}

 In summary, we studied the transmission properties of an infinitely long one dimensional lattice carrying a dimer in the center modeled by a saturable cubic-quintic discrete nonlinear Schr\"{o}dinger equation. The saturated cubic-quintic DNLS dimer was tested for asymmetric transmission of the input signal. With three possible ways to break the parity symmetry, we showed that transmission is more susceptible when broken symmetry corresponds to different on-site potentials. We also showed that if the lattice symmetry is broken by means of different cubic/quintic nonlinear parameters, the system supports better transmission at higher asymmetry. The rectifying action was computed to characterize the diode-like transmission and we reported that the rectifying action is reversed with regard to the transmission of right and left moving signal for the two types of symmetry breaking mechanisms (on-site asymmetric potential and/or on-site asymmetric nonlinearity). Further, we unveiled that, under the threefold broken symmetry, the diode-like action can be tuned to be \textit{positive} or \textit{negative}. Finally, the dynamics of a Gaussian wave-packet was considered numerically for the cases of asymmetry due to on-site potentials and the threefold asymmetry. The wave-packet scattering analysis confirms that the diode-action is reversed under the threefold asymmetry case. The above results demonstrate that the simultaneous control of the asymmetries on the linear and nonlinear parameters of a dimer defect can allow for an efficient diode-like action is specific spectral regions featuring both high transmission and large rectifying factor. We believe that the phenomenology predicted in our work can be experimentally probed, considering that several aspects of non-Hermitian and high-order nonlinear contributions have been unveiled in recent experiments regarding optical pulse propagation on laser-induced atomic gratings\cite{nonH1,nonH2,comp1,comp2}.  Efforts along this direction would bring valuable new insights to this exciting subject area.

\section*{Acknowledgments}

This work was supported by the ICT R \& D program of MSIT/IITP (1711073835: Reliable crypto-system standards and core technology development for secure quantum key distribution network) and GRI grant funded by GIST in 2018. MLL acknowledges the financial support from CNPq, CAPES, and FINEP (Federal Brazilian Agencies), and FAPEAL(Alagoas State Agency).

\section*{Competing Interests}

The authors declare no competing interests.

\section*{Author Contributions}

 M.A.W conceived the presented idea. M.A.W developed the theory, performed the analytical calculations and then the computations to produce the graphical results.
M.L.L verified the analytical results and analysed the graphical results. B.S.H reviewed the manuscript critically for important intellectual content, encouraged M.A.W and
supervised the findings of this work. All authors discussed the results and contributed to the final manuscript.

%\expandafter\ifx\csname natexlab\endcsname\relax\def\natexlab#1{#1}\fi
%\expandafter\ifx\csname bibnamefont\endcsname\relax
%  \def\bibnamefont#1{#1}\fi
%\expandafter\ifx\csname bibfnamefont\endcsname\relax
%  \def\bibfnamefont#1{#1}\fi
%\expandafter\ifx\csname citenamefont\endcsname\relax
%  \def\citenamefont#1{#1}\fi
%\expandafter\ifx\csname url\endcsname\relax
%  \def\url#1{\texttt{#1}}\fi
%\expandafter\ifx\csname urlprefix\endcsname\relax\def\urlprefix{URL }\fi
%\providecommand{\bibinfo}[2]{#2}
%\providecommand{\eprint}[2][]{\url{#2}}

\end{document}